# DESIGN OF MEBT FOR THE PROJECT X INJECTOR EXPERIMENT AT FERMILAB

A. Shemyakin[#], C. Baffes, A. Chen, Y. Eidelman, B. Hanna, V. Lebedev, S. Nagaitsev, J.-F. Ostiguy, R. Pasquinelli, D. Peterson, L. Prost, G. Saewert, V. Scarpine, B. Shteynas, N. Solyak, D. Sun, M. Wendt, V. Yakovlev, Fermilab*, Batavia, IL 60510, USA
T. Tang, SLAC, Menlo Park, CA 94025

*Abstract*

The Project X Injector Experiment (PXIE) [1], a test bed for the Project X front end, will be completed at Fermilab at FY12-16. One of the challenging goals of PXIE is demonstration of the capability to form a 1 mA H- beam with an arbitrary selected bunch pattern from the initially 5 mA 162.5 MHz CW train. The bunch selection will be made in the Medium Energy Beam Transport (MEBT) at 2.1 MeV by diverting undesired bunches to an absorber. This paper presents the MEBT scheme and describes development of its elements, including the kickers and absorber.

## REQUIREMENTS

The PXIE MEBT will be a ~10 m beam line between RFQ and the first Half-Wave Resonator cryomodule (HWR). It should form the required bunch pattern, match the optical functions between RFQ and SRF, include tools to measure properties of the beam coming out of RFQ and coming to SRF, and clean transverse halo particles while transporting the bunches selected for the following acceleration with a low emittance dilution and a low beam loss. The main MEBT requirements are listed in Table 1.

Table 1. The main MEBT functional requirements

| Parameter | Value | Unit |
|---|---|---|
| Beam kinetic energy | 2.1 +/-1% | **MeV** |
| Input frequency of bunches | 162.5 | MHz |
| Nominal input beam current | 5 | mA |
| Beam current operating range | 1-10 | mA |
| Nominal output beam current | 1 | mA |
| Relative residual charge of removed bunches | < $10^{-4}$ | |
| Beam loss of pass through bunches | < 5% | |
| Nominal transverse emittance | < 0.27 | μm |
| Nominal longitudinal emittance | 0.8 | eV·μs |
| Relative emittances increase | <10% | |

## FOCUSING SCHEME

Transverse focusing in the MEBT is provided mainly by equidistantly placed quadrupole triplets (Fig.1); the only exception is two doublets at the MEBT upstream end. Below the regions between neighboring triplets or doublets are referred to as MEBT sections. These sections are represented in Fig. 2 by rectangles color-coded according to their main function. The regular period is 1140 mm, which leaves 650 mm (flange-to-flange) space for equipment (350 mm in the section #0).

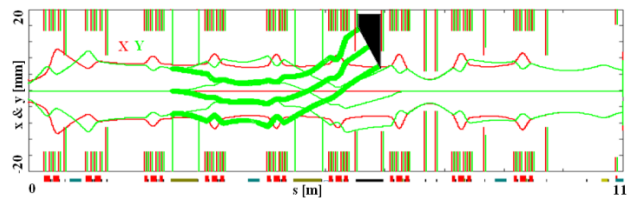

Figure 1: Scheme of MEBT optics [2] and the beam envelope. The thin lines are the central trajectory and 3σ envelope ($\varepsilon_{rms\_n}$=0.25 mm mrad) of the passing beam, and the thick lines are the Y envelope of the chopped-out beam. Red- quadrupoles, blue- bunching cavities.

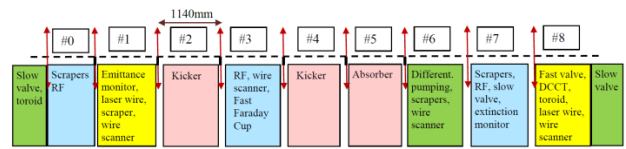

Figure 2: MEBT structure. Red - chopping system, blue- RF, yellow – diagnostics, green – vacuum.

Longitudinally the beam is focused (Fig.3) by three bunching cavities (Fig.4).

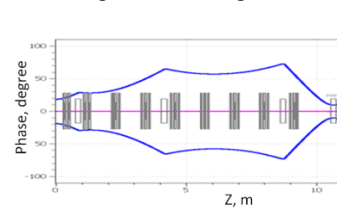 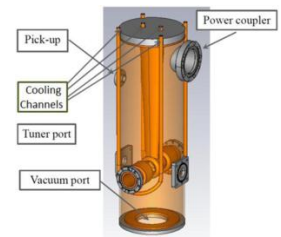

Figure 3: 3-σ bunch length through MEBT.  Figure 4: Conceptual design of the MEBT bunching cavity.

## CHOPPING SYSTEM

The undesired beam bunches will be removed in the MEBT by a chopping system consisting of two identical kickers separated by 180º transverse phase advance and an absorber at 90º from the last kicker. In the broadband, travelling-wave kickers [3], the transverse electric field propagates through their structures with the phase velocity equal to the speed of H- ions, 20 mm/ns.

Depending on polarity, bunches are either kicked toward the absorber surface or on a pass through path. In either case, the voltage on opposites plates is +/-250V with respect to ground. The gap between the kicker plates is 16 mm. Protection electrodes with 13 mm gap are installed on both sides of each kicker. To prevent extensive irradiation resulting from beam mismatch, an elevated electrode current will trigger a beam interruption. Because of critical importance of the kicker performance, presently two versions of the kicker are been developed.

In the first version having 50 Ohm characteristic impedance of the kicker structure, the beam is deflected by voltage applied to planar electrodes connected in vacuum by coaxial cables with the length providing necessary delays (Fig. 5). Each half of the ~0.5 m kicker has 25 copper plates with dimensions of 14.6 x 50 x 1 mm. The power loss induced by the electromagnetic wave (~100 W) and beam (~40W) is removed through the Teflon-insulated cables clamped to a water-cooled structure.

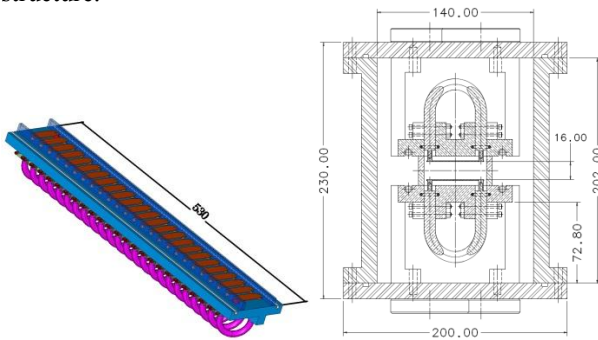

Figure 5: Mechanical schematic of the 50-Ohm kicker structure and the cross section of the kicker assembly.

The structure elements were simulated; a short mock-up was successfully measured; and presently a full-scale prototype is being designed.

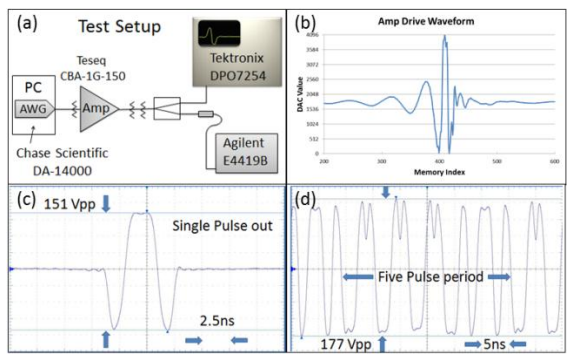

Figure 6: Test of the CBA 1G-150 amplifier with pre-distortion. (a) scheme of the test; (b) pre-distorted input signal and (c) corresponding output signal for a single pulse; (d) output for a CW pattern, corresponding to removal of four consecutive bunches followed by a one-bunch passage.

The kicker will be driven by a 1-kW linear amplifier. To decrease the lower frequency content of the output signal, the 6.15ns pulse affecting a single bunch is formed as shown in Fig.6c. Effects of a significant low frequency phase distortion in the commercially available amplifiers will be corrected by pre-distorting the driving signal generated according to an algorithm that has been successfully tested with a similar amplifier [4] of lower (150W) power (Fig.6).

The second version of the kicker has 200 Ohm impedance and consists of two helical structures (Fig. 7). Each helix is a flat wire wound with the 8.5mm helix pitch around a 28.6 mm OD copper grounded tube. The wire is suspended 4.8 mm above the tube surface by four ceramic spacers. The deflecting field is formed by flat copper 5.6x20x0.5 mm electrodes soldered to each turn of the helix.

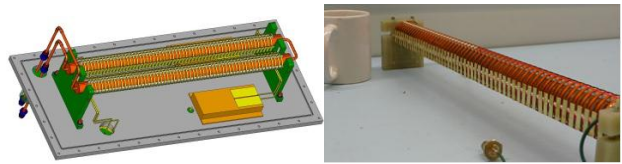

Figure 7: Mechanical drawing of 200 Ohm dual-helix kicker and photo of a single 200 Ohm helix prototype.

Simulations of the kicker structure agree well with prototype measurements (Fig. 8). Distortion during the propagation through the 200 Ohm helix is within specifications.

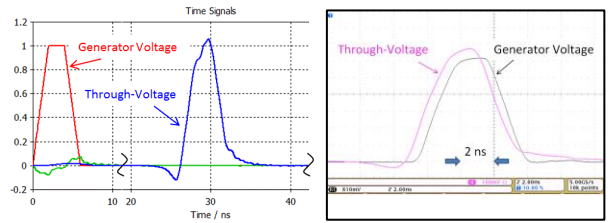

Figure 8: Simulation results of the helix kicker (left) and the prototype response (right). The traces at the right plot are time shifted.

The 200 Ohm kickers are intended to be driven by broadband, DC coupled switches in the push-pull configuration, which are under development [3].

The MEBT absorber should withstand irradiation by a 21 kW H- beam with the nominal rms beam radius of ~2mm. The design shown in Fig. 9 ([5]) addresses concerns about the heat load and resulting mechanical stresses as well as possible blistering issues.

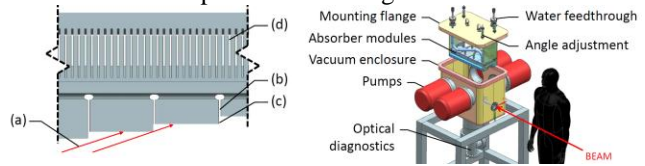

Figure 9: Conceptual design of the MEBT absorber. Left: side view of absorber showing (a) beam incident on surface, (b) axial stress relief slits, (c) shadowing step increment (magnitude exaggerated), (d) 300µm wide by 1mm pitch water cooling channels. Horizontal scale exaggerated. Right: exploded view.

To decrease the power density, the beam is directed to the 40-cm long absorber at a 29mrad grazing angle. The absorber is made of the molybdenum alloy TZM and is cooled by water flowing through transverse micro-channels. The axial stress is relieved by transverse slits. Thermal and stress analysis showed feasibility of the design. A 10-cm long prototype is being manufactured, and its thermal properties will be tested with an electron beam.

## OTHER SUBSYSTEMS

Several other subsystems are at different stages of development: quadrupoles, scrapers, vacuum system, and diagnostics.

The triplet assembly, which will consist of a central (larger) and two smaller quadrupoles, a pair of dipole correctors, and a button BPM mounted between the quadrupoles, is under technical design.

The scraping system concept assumes installation of 16 scrapers grouped in 4 sets with two X and two Y scarpers per set. Each scraper is an electrically isolated, ~100W-rated plate precisely movable across half of the aperture. The scrapers will be used for several purposes:
1. For beam halo measurements and removal
2. To protect the downstream equipment from beam losses. A high-loss signal from a scraper will trigger turning the beam turn-off.
3. As an axillary beam density distribution diagnostics (in a pulse mode)
4. To form a pencil H- beam for measurements downstream (in a pulse mode)

One pair of scraper sets is to be installed at the beginning of the MEBT to protect the MEBT elements (first of all, kickers), and the second pair is located after the absorber as a protection for cryomodules. The sets in each pair are separated by ~90º of the phase advance.

The vacuum system design is at a conceptual stage. The vacuum requirements are determined by the electron detachment in the H- beam and by the necessity to have a low gas flow into the HWR cryomodule. Obviously, the electron detachment results in a loss of H- beam intensity; an additional restrictive effect is a flux of created neutral hydrogen atoms that may reach the SRF cavities. For the design of the vacuum system, we adopted the requirement of keeping the integral of the pressure over the length of the MEBT to be below $1\cdot10^{-6}$ Torr·m, which corresponds to loosing ~$10^{-4}$ in the beam intensity and adding an additional ~0.1 W heat load to SRF by the neutrals.

The gas flow from the room-temperature MEBT to the 2K HWR cryomodule causes a gas deposition on the cryogenic surfaces. While there are no exhaustive experiments or a cohesive theory on the subject, there are indications in literature that such coverage at the surface of the cryogenic cavities degrades their performance. On the other hand, the closest analogue, a HWR in the Argonne National lab, successfully operates at the pressure of $5\cdot10^{-8}$ Torr immediately upstream of the cryomodule, though with the residual gas composed primarily by water [6]. To be on a safe side, we aim to keep the pressure upstream of PXIE's HWR at or below of $1\cdot10^{-9}$ Torr ($H_2$).

The main source of the vacuum load in the MEBT is expected to be a flow of hydrogen created from recombination of protons at the absorber. The worst scenario of 10 mA average absorber current corresponds to ~1 mTorr·l/s of the hydrogen load. To maintain a high vacuum near the SRF, first, the absorber is placed into a large box with effective pumping speed of 2500 l/s by turbo pumps and, second, the absorber is followed by a differential pumping section (#6 in Fig.2), where the vacuum chambers are separated by a 10mm ID, 200 mm long pipe. The calculated pressure profile is shown in Fig.10. Low particulate vacuum practices will be applied to the portion of the MEBT downstream of the section #6 to prevent degradation of the SRF cavities performance.

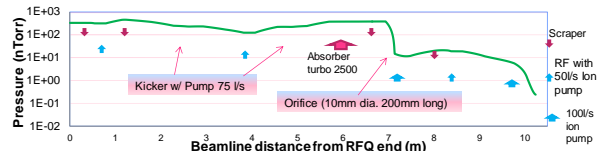

Figure 10: The pressure profile in the MEBT calculated for 10mA beam at the absorber.

The main purpose of two yellow-colored sections in Fig.2, #1 and #8, is to house beam diagnostics. A fully equipped MEBT diagnostics should allow characterizing the beam coming in and out of MEBT as well as its optics. The diagnostics are listed in Fig.2, and their detail description can be found in Ref. [7].

## CONCLUSION AND PLANS

The proposed scheme of the MEBT contains all elements required for shaping and characterizing the beam coming to the Project X SRF linac. The extensive test of the MEBT operation is planned to be carried out in PXIE. The FY2013 schedule includes manufacturing and testing prototypes for the absorber, kicker, quadrupoles, and BPM.

## ACKNOWLEDGMENT

The authors acknowledge multiple useful discussions of an alternative MEBT scheme with J. Staples.

## REFERENCES

[1] S. Holmes et al., Proc. of IPAC'12, New Orleans, USA, May 20 - 25, 2012, THPPP090
[2] V. Lebedev et al., ibid, THPPP057
[3] V. Lebedev et al., ibid, WEPPD078
[4] http://www.teseq.us/com/en/products_solutions/emc_radio_frequency/power_amplifiers/CBA_1G-150_e.pdf
[5] C. Baffes et al., Proc. of IPAC'12, New Orleans, USA, May 20 - 25, 2012, WEPPD035
[6] P. Ostroumov, G. Zinkann, private communication, 2012
[7] V. E. Scarpine et al., Proc. of IPAC'12, New Orleans, USA, May 20 - 25, 2012, MOPPR072